# Heat capacity anomaly near magnetic phase transition in GaFeO$_3$

Maria Biernacka[1], Marek Szafrański[2], Katarzyna Recko[1], Wojciech Olszewski[1], Dariusz Satuła[1], and Krzysztof R. Szymański[1]

[1]Faculty of Physics, University of Bialystok, K. Ciołkowskiego 1L, 15–245 Białystok, Poland.

[2]Faculty of Physics, Adam Mickiewicz University, Umultowska 85, 61-614 Poznań, Poland

Abstract

A heat capacity anomaly in the vicinity of magnetic transition was observed in single crystal GaFeO$_3$ for the first time. The crystals grown along the expected electric polarization direction were characterized by X-ray diffraction, Mössbauer and magnetic measurements. Two separate data analyses of heat capacity showed similar results within the experimental accuracy. Critical exponent, $\alpha = -0.2(1)$, and amplitude ratios of positive and negative reduced temperatures near the critical point, $A_+/A_- = 1.2(1)$, are consistent with behavior of a 3-D Heisenberg ferromagnet.

## 1. Introduction

GaFeO$_3$ compound have attracted significant attention since the discovery of their multiferroic properties [1], linear magnetoelectric effect [2], large magnetic anisotropy [3], and magnetization [4]. Particularly important for application is their potential to control magnetic transition temperatures near room temperature by changing the Fe/Ge ratio [5, 6]. The compound crystallizes in the orthorhombic structure of space group No. 33 [7, 8, 9, 10, 11], *Pna*2$_1$ or *Pc*2$_1$*n*, depending on the choice of the crystallographic axes.

For applications, it is very important that magnetism and ferroelectricity simultaneously occur in a single phase at room temperature (RT). This was demonstrated on orthorhombic GaFeO$_3$ thin films with an excess of iron, grown by pulsed laser deposition technique [12]. RT ferroelectricity was observed on epitaxial thin films [13, 14]. Leakage currents hinder the observation of the electrical polarization switching in GaFeO$_3$, and to date it has only been demonstrated on undoped [13, 15] or Mg-doped GaFeO$_3$ films [16]. Katayama et al. [17] have shown that Cr-substituted GaFeO$_3$ epitaxial films exhibit simultaneous in-plane ferrimagnetism and out-of-plane ferroelectricity. In these films, in contrast to bulk GaFeO$_3$, some Fe ions were located at tetrahedrally coordinated Ga sites. The authors also observed that the shape of the hysteresis curves for electrical polarization switching in Cr-substituted samples was inconsistent with earlier proposed switching mechanisms [13, 18, 19]. In this context, it is astonishing that no switching polarization mechanism, consistent with theoretical prediction, has been demonstrated on bulk GaFeO$_3$ single crystals. It is also unexpected that such an important thermodynamic property as heat capacity has only been published for high temperatures [20]. However, the heat capacity anomaly at magnetic transition temperature has not yet been experimentally detected [21].

In this paper, we present the results of heat capacity measurements near the magnetic transition temperature on GaFeO$_3$ single crystals, grown successfully by our group. Relevant structural and magnetic characteristics are also reported.

## 2. Sample preparation

GaFeO$_3$ single crystals were grown using the optical floating zone crystal growth technique following previous approaches [8, 10, 22]. The starting materials were Fe$_2$O$_3$ (99.999%, Acros Organics) and Ga$_2$O$_3$ (99.99%, Sigma Aldrich) powders. Stoichiometric amounts of the oxides were thoroughly ground together, pre-calcined in a horizontal furnace tube at 1100°C for 10 h and calcined at 1250°C over next 10 h in the air, with intermediate regrinding. Then the powder was loaded into the rubber tubes and compacted using a hydraulic press at a pressure of 70 MPa. The obtained rods were sintered for 72 h in air at 1385°C in a vertical furnace tube and then cooled to room temperature at a rate of 1°C/min. Single crystals growth was performed with a four-mirror optical floating zone furnace (FZ-T-4000-H, Crystal Systems Corp. Japan, lamps power 4x300 W) in a pure oxygen. To suppress incongruent decomposition [23], the growth process was carried out under pressure of 9.0-9.2 bars with an oxygen flow rate of 0.4 l/min. Crystals were grown at the rate of 3-5 mm/h, with feed and seed rods rotated at 15 rpm in the opposite direction.

The crystal growth was designed to have electrical polarization axis parallel to the crystal growth direction. To achieve this goal, a seed rod with appropriate orientation was prepared. First, a polycrystalline rod was used as a seed and a necking technique was applied during growth. In this way, a single crystal was obtained by spontaneous nucleation. The crystal was next cut to form a pyramid a few millimeters high and appropriately oriented. The base of the pyramid was attached with polyvinyl acetate binder to the top of the flat polycrystalline rod. The rod with the pyramid on the top was sintered at 1380°C in the air for 48 h, and slowly cooled to RT, resulting in a sufficiently strong joint.

The orientation of the grown crystal was checked for a slice cut off from the grown rod, by single crystal X-ray diffraction. The measurements revealed that the direction of the crystal growth was inclined by 12° from the polar axis, that is, to the direction parallel to the longest unit cell parameter.

3. **Experimental details**

The single crystal diffraction experiments were performed on an Oxford Diffraction Gemini A Ultra diffractometer operating with graphite monochromated Mo$K_\alpha$ radiation ($\lambda$ = 0.71073 Å). CrysAlisPro software [24] was used for data collection and processing. The crystal structure was solved with direct methods using SHELXS-97 and refined by the full-matrix least-squares method on all intensity data with SHELXL-97 [25].

Structural characterization of the powder obtained by crushing part of a single crystal was carried out using an Empyrean PANalytical powder diffractometer utilizing Mo$K_\alpha$ radiation ($\lambda$ = 0.7093187 Å in Bragg-Brentano geometry). The data were acquired in a $2\theta$ range from 5° to 75°, using a step value of 0.026°. The diffraction patterns were analyzed by the Rietveld-type profile refinement method using FullProf software [26].

Mössbauer measurements were performed at room temperature using a spectrometer operating in a constant acceleration mode with a $^{57}$Co source in an Rh matrix. The velocity scale was calibrated using α-Fe standard foil at room temperature. The spectrum was analyzed by commercially available NORMOS package [27] using the transmission integral approach.

Magnetic measurements were performed with a vibrating sample magnetometer (Cryogenic Limited) in a temperature range of 2-300 K, and a magnetic field strength up to 9 T. Magnetization was measured at a temperature of 2 K, as a function of the magnetic field.

Heat capacity measurements were performed with an AC micro-calorimeter (Cryogenic Limited). The sample, weighing between 1 and 2 µg, was mounted using a small amount of grease (ApiezonN) onto a silicon nitride membrane with heathers and temperature sensors embedded. An AC current was applied to the heather, and the resulting temperature oscillations were measured. AC current amplitude was adjusted to the range of the temperature oscillations of about 0.5 K. Continuous cooling and warming curves with a 0.2 K/min. ramp were taken, in the temperature range 190-210 K. The heat capacity of ApiezonN and the sample holder were measured separately, and the resulting measurements did not reveal any signs of heat capacity anomalies. The background signal value was as high as 50% of the total signal from the sample with holder and grease.

## 4. Results

### a) Structural characterization

The quality of the single crystals was checked by X-ray diffraction, using a small piece cut from the large crystal. The structure was solved in space group $Pc2_1n$, and the refinement of the structural model was initially carried out assuming the stoichiometric ratio of Ga and Fe atoms, as well as their complete order. Relatively high values of $R$-factors ($R_1/wR_2$ ($I > 2\sigma_I$) = 0.0315/0.0910 and $R_1/wR_2$ (all data) = 0.0322/0.0917), the high residual electron peak andelectron hole on the difference Fourier map (1.584 and −1.839 e Å$^{-3}$, respectively), and the odd thermal parameters of some atoms suggested a possible disorder and/or a non-stoichiometry of the crystal. Therefore, in the next step the disordered model was applied with the use of the Mössbauer spectroscopy results. This spectroscopy is particularly useful in the studies of GaFeO$_3$ because the Mössbauer spectrum reveals the chemical disorder of the crystal. The Mössbauer spectrum of the powder GaFeO$_3$ sample measured at room temperature is shown in Fig. 1. Three doublets, corresponding to different iron crystallographic positions, were fitted to the spectrum. The obtained values of the doublets hyperfine field parameters were fully consistent with earlier reported measurements on single crystals [28], [29] and polycrystalline samples [30], [31], [32]. The relative contributions of the three doublets were found to be 39%, 35% and 26%; these values were used to determine the site occupancy factors listed in Table 1, column 2. The Ga(1) site is occupied solely by Ga atoms, while sites Fe(1), Ga(2) and Fe(2) are shared among Ga and Fe atoms. The coordinates of the atoms occupying the same sites were refined without any constraints (Table 1, columns 3 to 5). For the disordered structure, the reliability $R$-factors decreased substantially (Table 1), and the residual electron peak and electron hole values improved to 0.512 and −0.517 e Å$^{-3}$, respectively, indicating a substantial improvement of the structural model.

Another independent approach of structural characterization was performed on the powder obtained by crushing a single crystal and using combined Mössbauer and X-ray diffraction techniques. The site occupancies listed in the second column of Table 1 were used to obtain structural parameters within the FullProf application, based on the Rietveld structure refinement routine (Table 1, columns 6 to 8).

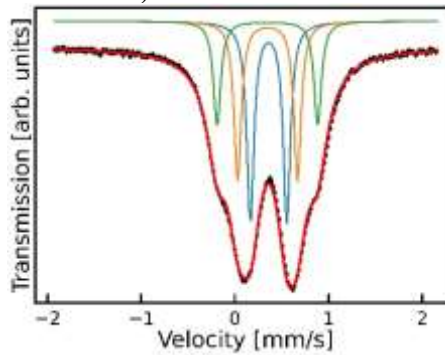

Fig. 1. Mössbauer spectrum of powdered single crystal GaFeO$_3$ sample measured at room temperature. Spectrum was fitted by three components shown at the upper part of the figure. Red line shows result of the best fits within the transmission integral approximation.

Table 1. Selected refinement details of GaFeO$_3$ structure (*Pc*2$_1$*n* space group) obtained from the RT single crystal (col. 3-5) and powder X-ray diffraction data (col. 6-8).

| | | single crystal | | | powder | | |
|---|---|---|---|---|---|---|---|
| 1 | 2 | 3 | 4 | 5 | 6 | 7 | 8 |
| Site | Fe/Ga [%] | $x$ | $y$ | $z$ | $x$ | $y$ | $z$ |
| Ga(1) | 0/100 | 0.1522(1) | -0.0025(1) | 0.1764(1) | 0.1497(9) | 0 | 0.1760(12) |
| Ga(2) | 52/48 | 0.1614(4) | 0.3041(5) | 0.8176(7) | 0.1571(9) | 0.3059(9) | 0.7995(10) |
| | | 0.1596(3) | 0.3071(4) | 0.8070(6) | | | |
| Fe(1) | 70/30 | 0.0305(2) | 0.7956(4) | 0.6744(5) | 0.0318(6) | 0.7941(12) | 0.6793(12) |
| | | 0.0373(4) | 0.7930(6) | 0.6807(7) | | | |
| Fe(2) | 78/22 | 0.1540(3) | 0.5811(2) | 0.1875(4) | 0.1566(12) | 0.5827(7) | 0.1925(19) |
| | | 0.1467(7) | 0.5841(7) | 0.1993(11) | | | |
| O(1) | | 0.3236(2) | 0.4246(2) | 0.9796(5) | 0.3340(31) | 0.4187(23) | 0.9383(49) |
| O(2) | | 0.4887(2) | 0.4303(2) | 0.5184(4) | 0.5110(29) | 0.4159(45) | 0.4858(54) |
| O(3) | Fixed to | 0.9992(2) | 0.1998(3) | 0.6530(3) | 0.9996(26) | 0.2144(25) | 0.6628(67) |
| O(4) | 100 | 0.1597(2) | 0.1950(2) | 0.1558(3) | 0.1676(36) | 0.1815(19) | 0.2036(61) |
| O(5) | | 0.1700(2) | 0.6705(3) | 0.8466(3) | 0.1694(34) | 0.6663(24) | 0.8501(57) |
| O(6) | | 0.1697(2) | 0.9373(2) | 0.5159(4) | 0.1684(30) | 0.9521(28) | 0.5054(58) |
| Cell parameters [Å] | | $a$ | $b$ | $c$ | $a$ | $b$ | $c$ |
| | | 8.7304(1) | 9.3761(2) | 5.0759(1) | 8.7320(8) | 9.3762(8) | 5.0757(4) |
| Reliability factors | | $R_{int}$ | $R_1/wR_2$ ($I>2\sigma_I$) | $R_1/wR_2$ (all data) | Bragg $R$ | $Rf$ | $\chi^2$ |
| | | 0.0298 | 0.0150/0.0408 | 0.0156/0.0411 | 0.093 | 0.076 | 7.06 |

b) **Magnetic measurements**

The sample for magnetization measurements was prepared in a cuboid shape with edges parallel to the main crystal directions. Magnetization versus temperature was measured at an external magnetic field of 0.05 T, as shown in Fig. 2. The results of magnetic measurements agree with previous research [8, 22] showing large anisotropy with an easy magnetic axis parallel to the shortest edge *c* of an elementary cell. The phase transition temperature obtained from an inflection point of a field-cooling (FC) curve for a magnetic field of 0.05T applied in *c* direction was 206(5) K. The *M*(*H*) measurements (inset in Fig. 2) yield a spontaneous magnetization value of 0.74 $\mu_B$/ Fe at 2 K, which is slightly higher than the value of 0.67 $\mu_B$/Fe at 5 K reported by Arima [8]. The coercivity is slightly lower than that reported by others [8, 21], very likely because of a small excess of iron.

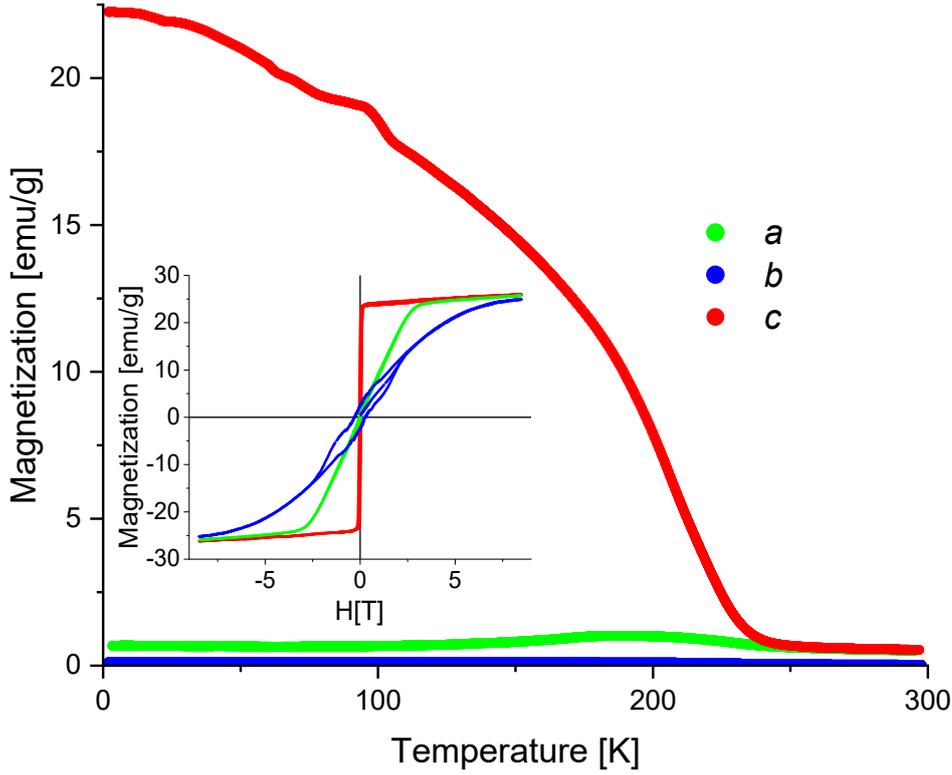

Fig. 2. FC magnetization curves under external magnetic field of 0.05 T. The inset shows magnetization as a function of magnetic field at temperature of 2 K. $M(T)$ and $M(H)$ curves were measured with a magnetic field aligned parallel to particular axes $a$, $b$, $c$ of a single crystal. Direction $c$ is parallel to the shortest parameter of the unit cell of $GaFeO_3$, and it is an easy magnetization axis ($Pc2_1n$ space group).

### c) Heat capacity measurements

Heat capacity measurements were carried out with a zero applied magnetic field and a magnetic field of 1 T (Fig. 3). The singular part of heat capacity $C$ was obtained by subtracting from the measured total heat capacity the non-singular part of heat capacity $C_n$, where $C_n$ consists of background (holder+ApiezonN) and non-singular components of the heat capacity of $GaFeO_3$ lattice and electron contribution. We observed that the simple linear temperature function did not describe the $C_n$ data well; therefore, it was modeled by second- or third-order polynomial (the red line in Fig. 3) fitted to the heat capacity data in temperature ranges far from $T_C$ (175-190 K, 207-215 K). The singular part $C$ measured in the zero magnetic field has a clearly asymmetric cusp (Fig. 3, bottom right-hand side inset). External field of 1 T apparently suppresses the singularity behavior characteristic for second-order phase transition.

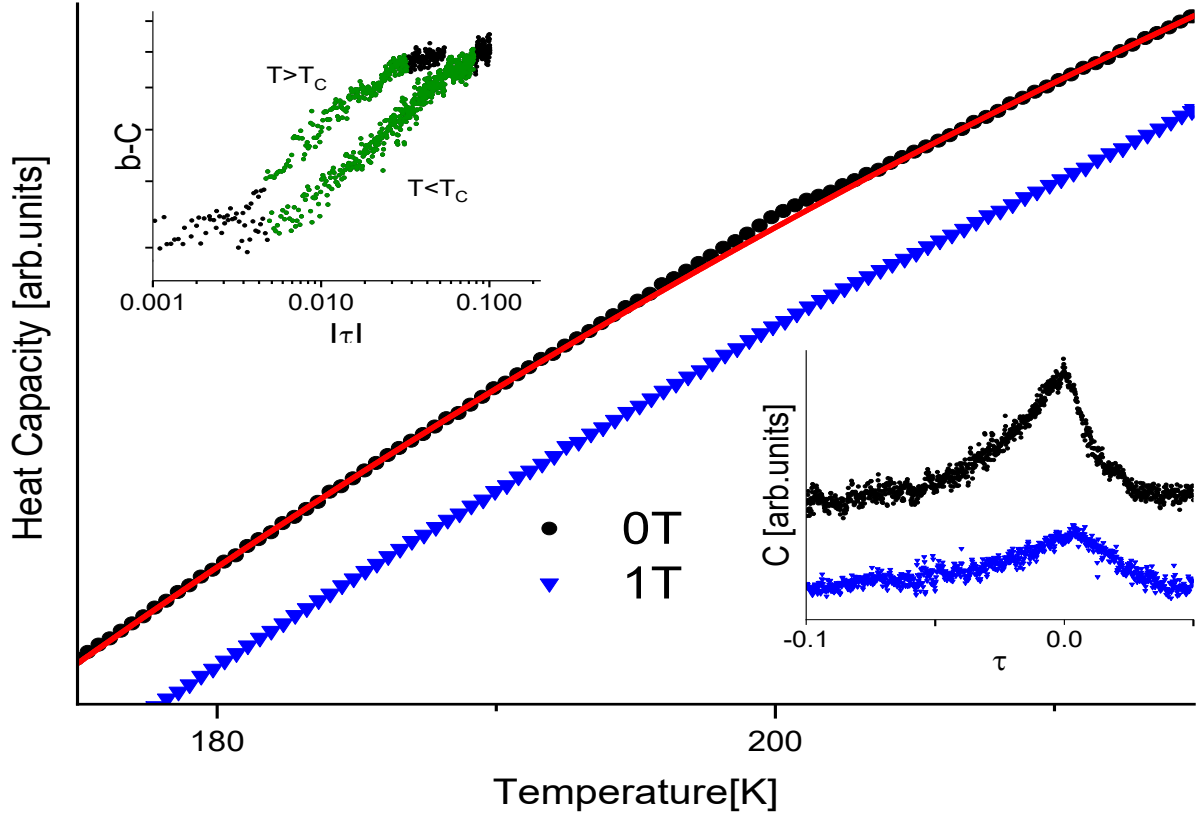

Fig. 3. Specific heat of single crystal GaFeO$_3$ measured in external magnetic field strengths of zero and 1 T. The red line shows the non-singular part, the singular part (bottom right-hand side inset) and the log-log plot of the singular part of the specific heat (upper left-hand side inset).

Although the signal-to-noise ratio for our data is poor, it is sufficient to indicate the sign of the critical exponent $\alpha$ of the singular dependence of heat capacity $C$ on the reduced temperature $\tau$: $C \sim |\tau|^{-\alpha}$, where $\tau = T/T_C - 1$ and $T_C$ are the transition temperatures. By choosing appropriate values of the two parameters $T_C$ and $b$, we plot $b - C$ vs $|\tau|$ in logarithmic scales. There is a relatively wide region of reduced temperatures for which data points are located on two clearly separated straight lines (see green points, Fig. 3, upper left-hand side inset). However, the range of reduced temperatures for which data are located on straight lines, as well as the choice of the $T_C$ and $b$ parameters, are to some extents arbitrary and depend upon the details of the $C_n$ subtraction. Thus, fitting procedures were used for further analysis. Because the detected signal-to-noise ratio was weak, we attempted to use two approaches in order to minimize the number of free parameters.

In the first approach the singular part of heat capacity $C$ was quantitatively described by function [33], [34]

$$C(\tau) = \frac{A_\pm}{\alpha} |\tau|^{-\alpha} + b, \tag{1}$$

where $A_+$ is amplitude for $\tau > 0$, while $A_-$ is amplitude for $\tau < 0$. In these fits there were five free parameters: $T_c, \alpha, A_+, A_-, b$. The data with small values of $|\tau|$ were excluded because of peak rounding, following the widely accepted data treatments [33, 35]. By using the nonlinear least squares Levenberg-Marquardt method, we fitted the $C(\tau)$ function to the experimental data. An example is shown Fig. 4, and the relevant parameters are listed in Table 2, together with the details of the data handling.

The second approach is based on the fact that measurements are realized by temporal oscillatory variation of the heat load to the sample. Thus the temperature is constantly oscillating, with an approximate peak-to-peak value of about 0.5 K. One expects that this technical issue results in additional rounding of the transition. We have used the convolution of the Gaussian distribution $f(\tau)$ with variance $\sigma^2$ and the singular part (1) as a function describing the data:

$$C^*(\tau) = \int_{-\infty}^{-\infty} C(t)f(\tau - t)dt. \tag{2}$$

Explicit forms of functions $f(\tau)$ and $C^*(\tau)$ are given in the Appendix. The uncertainties of the parameters obtained in the fits were estimated as statistical error propagation, using the nonlinear least square method [36]. Example of the best fit is shown in Fig. 4 and the relevant parameters are listed in Table 2.

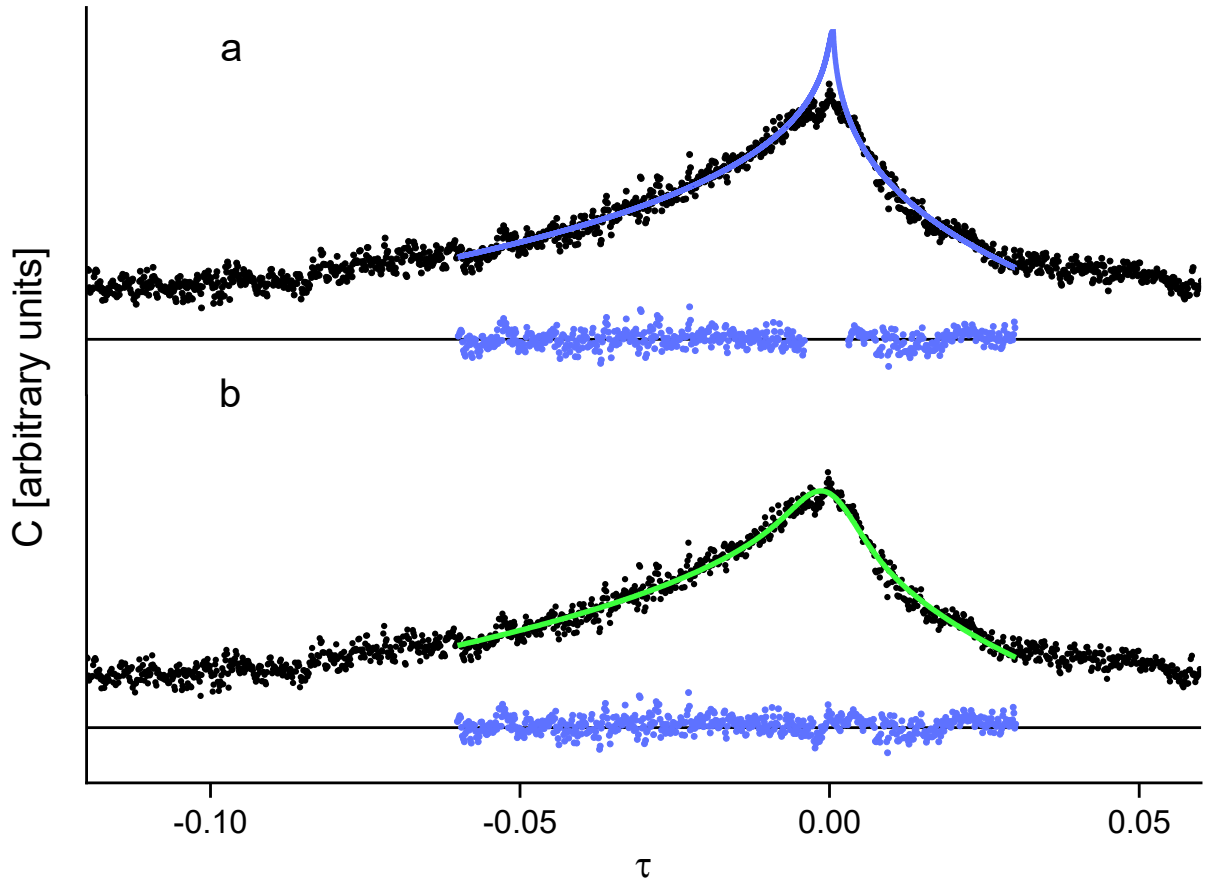

Fig. 4. Examples of the fit of *a)* $C(\tau)$ and *b)* $C^*(\tau)$ (2) functions to the singular part of data. Residues are shown to indicate the quality of the fits.

Table 2
The results of fitting: of function $C(\tau)$ to the data that are in the range $(\tau_1, \tau_2)$ but not in the range $(\tau_1^0, \tau_2^0)$, close to the critical region; and of function $C^*(\tau)$ to the data that are in the range $(\tau_1, \tau_2)$. Parameter $n$ is the order of the polynomial fitted to the heat capacity data far from the critical region. Symbols $\alpha, T_c, A_+, A_-, \sigma$ are parameters of functions $C(\tau)$ (1) and $C^*(\tau)$ (2).

| Fitted function | $\tau_1, \tau_2$ | $\tau_1^0, \tau_2^0$ | $n$ | $\alpha$ | $T_C$ [K] | $A_+/A_-$ | $\sigma$ [K] |
|---|---|---|---|---|---|---|---|
| $C$ | -0.06,0.03 | -0.004,0.003 | 2 | -0.28(3) | 200.53(3) | 1.3(2) | - |
| $C$ | -0.08,0.04 | -0.004,0.003 | 2 | -0.11(2) | 200.55(2) | 1.1(1) | - |
| $C$ | -0.06,0.03 | -0.004,0.003 | 3 | -0.25(3) | 200.57(3) | 1.3(2) | - |
| $C$ | -0.08,0.04 | -0.004,0.003 | 3 | -0.07(2) | 200.60(2) | 1.1(1) | - |
| $C^*$ | -0.06,0.03 | - | 2 | -0.32(9) | 200.51(8) | 1.3 (1) | 0.7(2) |
| $C^*$ | -0.08,0.04 | - | 2 | -0.22(6) | 200.50(6) | 1.23(6) | 0.6(1) |
| $C^*$ | -0.06,0.03 | - | 3 | -0.36(8) | 200.47(8) | 1.35(9) | 0.7(2) |
| $C^*$ | -0.08,0.04 | - | 3 | -0.16(6) | 200.43(8) | 1.14(6) | 0.9(1) |

## 5. Discussion and summary

The results of structural analysis of the Ga-Fe-O crystal grown in *c* direction are consistent with previously reported data for crystals grown in other unspecified directions. The Ga site with tetrahedral coordination is almost perfectly filled by Ga, while the three other sites exhibit chemical disorder. Structural parameters measured on pieces of single crystals (Table 1, columns 3 to 5) agree with those obtained for powders (Table 1, columns 6 to 8) and reported results [8, 9, 10, 11, 37]. The magnetic measurement data indicate a small excess of iron.

The different sample preparation methods used in the experiments, influencing mainly features of the Ga/Fe site disorder, result in different magnetic transition temperatures $T_c$. The $T_c$ measured by magnetization and heat capacity agree with the temperature reported for single crystals grown by the floating-zone method, that is, 200 K [8]. Because of larger chemical disorder introduced by high temperatures of the floating-zone process, however, these temperatures are lower than $T_c$ of crystals obtained by flux growth, namely, 255 K [38], 285 K [4], 260 K [1], 260 K [39], and 292 K [40].

The plot of $b - C$ vs $|\tau|$ clearly shows that in some regions of reduced temperature, data points are located on straight lines with positive slopes (inset, Fig. 3), thus the critical exponent $\alpha$ in the relation $C|\tau|^{-\alpha}$ is negative. The separation of the lines in the inset is also consistent with the asymmetry of the cusp shown (right inset), as well as the values $A_+/A_-$ presented in Table 2. Both methods of data treatment – fitting the $|\tau|^{-\alpha}$ function (1) to the data with narrow, critical range omitted; and fitting the convolution of the $|\tau|^{-\alpha}$ with Gaussian distribution (2) – produce consistent results. Moreover, the fitted values of the Gaussian distribution width (free parameter $\sigma$, Table 2) agree with the amplitude of the temperature oscillations characteristic for the measurement device. Use of function $C^*(\tau)$ results in some advantages and disadvantages, however. In contrast to fitting the $C(\tau)$ function, it is not necessary to exclude data close to the phase transition temperature. However, $C^*(\tau)$ requires one additional free parameter to be fitted, the width of the Gaussian distribution $\sigma$. We have derived the explicit shape of the convolution (2), so it is not necessary to perform numerical integration during the fitting.

The mean values of the parameters presented in Table 2 result in temperature $T_C = 200.5(1)$ K, $\alpha = -0.2(1)$ and $A_+/A_- = 1.2(1)$. The two last values are close to the theoretical predictions for 3D Heisenberg model, in which $\alpha = -0.14(6)$ [41], $\alpha = -0.1336(15)$, and $A_+/A_- = 1.56(4)$ [34].

Acknowledgments. This work was partly supported by the National Science Centre, Poland, under grant OPUS no 2018/31/B/ST3/00279 and by the Polish Government Plenipotentiary for JINR (Project no PWB/168-10/2021) and the Polish Ministry of Science and Higher Education under subsidy for maintaining the research potential of the Faculty of Physics, University of Bialystok.

## Appendix

The singular part of the heat capacity (1) can be rewritten to the form
$$C(\tau) = \alpha^{-1}\big(A_-\theta(-\tau) + A_+\theta(\tau)\big)|\tau|^{-\alpha} + b, \qquad (3)$$
where $\theta(\tau)$ is the Heaviside step function

$$\theta(\tau) = \begin{cases} 0, & \tau < 0 \\ 1/2, & \tau = 0. \\ 1, & \tau > 0 \end{cases} \quad (4)$$

Gaussian distribution $f(\tau)$ having variance $\sigma^2$,

$$f(\tau) = \frac{1}{\sqrt{2\pi}\sigma} \exp\left(-\frac{\tau^2}{2\sigma^2}\right), \quad (5)$$

convoluted with (3),

$$C^*(\tau) = \int_{-\infty}^{-\infty} C(t)f(\tau - t)dt + b, \quad (6)$$

can be calculated using the theorem on Fourier transform of convolution. The explicit form of $C^*(\tau)$ is:

$$C^*(\tau) = \frac{1}{\sqrt{\pi}2^{1+\alpha/2}\sigma^\alpha}(C_+ + C_-) + b, \quad (7)$$

$$C_+ = \frac{A_- + A_+}{\alpha}\Gamma\left(\frac{1-\alpha}{2}\right)M\left(\frac{\alpha}{2}, \frac{1}{2}, \frac{-\tau^2}{2\sigma^2}\right),$$

$$C_- = \frac{A_- - A_+}{\sqrt{2}\sigma}\tau\Gamma\left(\frac{-\alpha}{2}\right)M\left(\frac{\alpha+1}{2}, \frac{3}{2}, \frac{-\tau^2}{2\sigma^2}\right),$$

where $\Gamma$ is the Euler gamma function

$$\Gamma(z) = \int_0^\infty x^{z-1}e^{-x}dx, \quad (8)$$

and $M$ is Kummer's function of the first kind:

$$M(a,b,x) = 1 + \frac{a}{b}x + \frac{a(a+1)}{b(a+1)2!}x^2 + \frac{a(a+1)(a+2)}{b(b+1)(b+2)3!}x^3 + \cdots. \quad (9)$$